\documentclass[prl, twocolumn, showpacs, nofootinbib, amsmath,amssymb, floatfix, eqsecnum]{revtex4-1}
\pdfoutput=1
\usepackage{amsmath}
\usepackage{amssymb}
\usepackage{amsthm}
\usepackage{amsfonts}
\usepackage{comment}

\usepackage{graphicx}
\usepackage{color,framed}
\usepackage{hyperref}
\usepackage{times}
\usepackage{enumerate}
\usepackage{lipsum}
\usepackage{slashed}
\usepackage{url}
\usepackage{bbm}
\usepackage{chngcntr}
\counterwithout{equation}{section}

\hypersetup{
    colorlinks=true, 
    linktoc=all,     
    linkcolor=blue,  
}

\def \beq {\begin{equation}}
\def \eeq {\end{equation}}
\def \beqa {\begin{eqnarray}}
\def \eeqa {\end{eqnarray}}
\def \bseq {\begin{subequations}}
\def \eseq {\end{subequations}}

\begin{document}

\title{Volume-law to area-law entanglement transition in a non-unitary periodic Gaussian circuit}

\author{Etienne Granet and Carolyn Zhang}
\affiliation{Department of Physics, Kadanoff Center for Theoretical Physics, University of Chicago, Chicago, Illinois 60637,  USA}
\author{Henrik Dreyer}
\affiliation{Quantinuum, Leopoldstrasse 180, 80804 Munich, Germany}
\date{\today}

\begin{abstract}
We consider Gaussian quantum circuits that alternate unitary gates and post-selected weak measurements, with spatial translation symmetry and time periodicity. We show analytically that such models can host different kinds of measurement-induced phase transitions detected by entanglement entropy, by mapping the time evolution and weak measurements to M\"obius transformations. We demonstrate the existence of a log-law to area-law transition, as well as a volume-law to area-law transition. For the latter, we compute the critical exponent $\nu$ for the Hartley, von Neumann and R\'enyi entropies exactly.
\end{abstract}

\maketitle

\textbf{\emph{Introduction.}}---In recent years, there has been an immense amount of work on dynamical phase transitions driven by competition between unitary time evolution and projective measurements, called measurement-induced phase transitions (MIPT). It was found in Ref.~\onlinecite{skinner2019,li2018,li2019,chan2019} that, although generic unitary time evolution leads to volume-law entangled states in the long-time limit, interspersing the unitary evolution with a sufficient frequency of local measurements can stabilize area-law entangled steady states. These MIPTs were observed in other setups as well, and were shown to depend crucially on the type of unitaries and measurements applied\cite{szyniszewski2019entanglement,jian2021measurement,nahum2021measurement,buchhold2021effective,tang2020measurement,gopalakrishnan2021entanglement,turkeshi2020measurement,block2022measurement,agrawal2022entanglement,sang2021entanglement,szyniszewski2020universality,goto2020measurement,lunt2020measurement,botzung2021engineered,lopez2020mean,noel2022measurement,iaconis2020measurement,yang2022entanglement}. There is a subclass of circuits called Gaussian circuits that allow for analytical calculations because they only involve unitaries and measurements built out of fermion bilinears. Although the volume-law to area-law MIPT was observed for Haar random unitaries, Clifford random circuits and interacting Floquet circuits, it was found that in the original setup, Gaussian circuits can only produce sub-volume law entanglement states whenever there is a non-zero measurement rate \cite{cao2019,chen2020,nahum2020,alberton2021,jian2022,fidkowski2021,carollo2022}. In addition, analytical results exist only for the Hartley entropy \cite{skinner2019}, in the limit of infinite-dimensional spins \cite{bao2020theory,jian2020measurement}, or for dual-unitary circuits \cite{bertini2020scrambling,lu2021,klobas2021entanglement}, as well as in continuous-time free fermionic models \cite{cao2019,legal2022}. Hence, an analytically tractable circuit model for qubits displaying a MIPT for the von Neumann or R\'enyi entropies is still lacking. 

In this work, we analytically study Gaussian non-unitary circuits with spatial translation symmetry and discrete time translation symmetry. They consist of alternating Gaussian unitary gates and weak measurements, obtained by coupling the system to an ancilla, measuring the ancilla, and post-selecting \cite{supp}. We show that these non-unitary circuits can demonstrate MIPTs between different entanglement phases, as we tune the strength of the measurement, which is related to the coupling to the ancilla. Surprisingly, we show that for a particular choice of parameters, we can tune through a volume-law to area-law transition. We derive exact properties of the critical points, including the critical weak measurement amplitude and the correlation length exponent $\nu$ for the Hartley, von Neumann and R\'enyi entanglement entropies $S_0(\ell),S_1(\ell),S_m(\ell)$ ($m\geq 2$) between a subsystem of size $\ell$ and its complement. To our knowledge, this is the first example of a volume-law to area-law transition in a Gaussian non-unitary circuit, and of an analytical computation of critical exponents of von Neumann and R\'enyi entropies at a MIPT.
\begin{figure}[tb]
   \centering
   \includegraphics[width=.8\columnwidth]{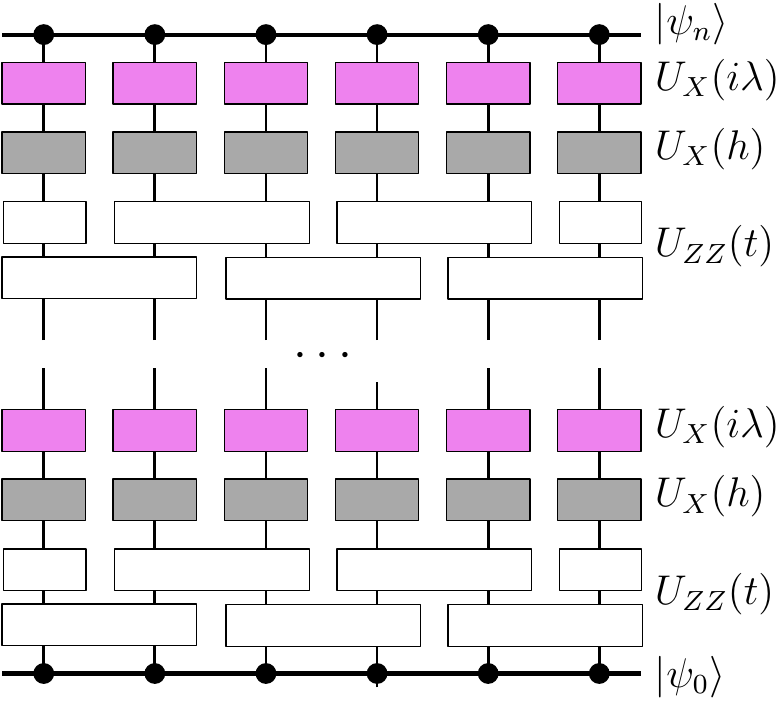} 
   \caption{A schematic of $n$ applications of the round described by (\ref{U}), with $p=1$.}
   \label{fig:fig1}
\end{figure}
   
\textbf{\emph{Setup.}}---We study non-unitary circuits built out of the following translation-invariant 1D layers: \cite{dreyer2021,granet2022}
\begin{align}
\begin{split}\label{layers}
U_{ZZ}(t)&=e^{-it\sum_{j=1}^L\sigma^z_j\sigma^z_{j+1}}\\
U_{YY}(t)&=e^{-it\sum_{j=1}^L\sigma^y_j\sigma^y_{j+1}}\\
U_{X}(t)&=e^{-it\sum_{j=1}^L\sigma^x_j}
\end{split}
\end{align}
where $\sigma^{x,y,z}_j$ are Pauli matrices, and periodic boundary conditions $L+1\equiv 1$ are assumed. These layers can be written as free fermion evolution using the standard Jordan-Wigner transformation $\sigma^x_j= 1-2c_j^\dagger c_j$ and $\sigma^z_j=(c_j+c_j^\dagger)\prod_{\ell=1}^{j-1}(1-2c_\ell^\dagger c_\ell)$, where $c_j^\dagger, c_j$ are fermion creation and annihilation operators. Non-unitarity is introduced by allowing $t$ to be complex, which corresponds to weak measurement \cite{supp}. 

\begin{figure}[tb]
   \centering
\includegraphics[width=0.95\columnwidth]{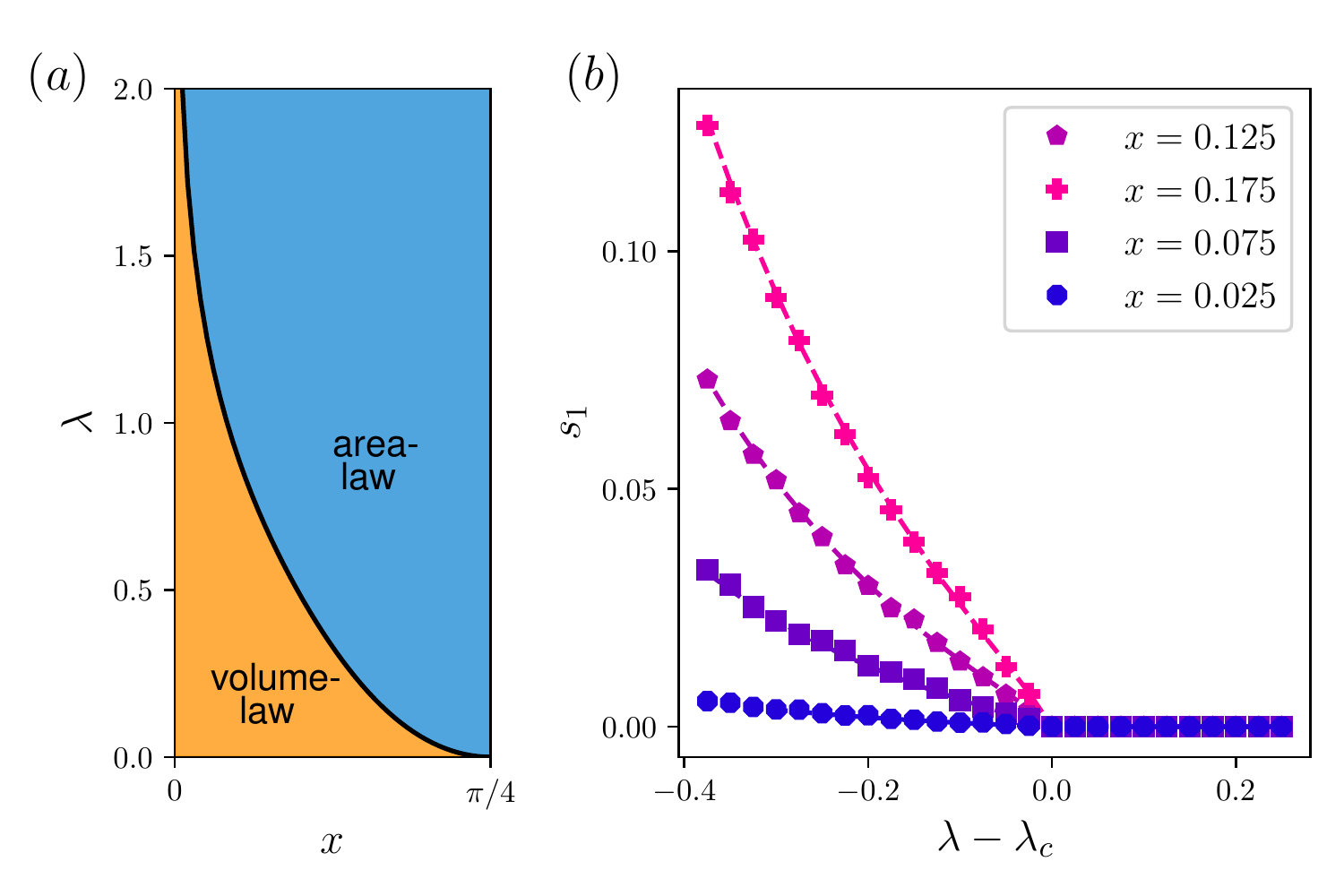} 
   \caption{The non-unitary circuit described in (\ref{twotimes}) gives a volume-law to area-law transition. (a) Phase diagram in the space of competing parameters: $x$ denotes the difference in duration of the two cycles~(\ref{definition_of_x}) and $\lambda$ is the measurement strength~(\ref{constt}). The phase boundary is given by~(\ref{lambdacvolume}).
   (b) Scaling of the von-Neumann entropy. Markers denote the slope of the best linear fits to the exact entropy $S_1(\ell) \sim s_1 \ell + b$ after 500 cycles on subsystem sizes up to $\ell = 100$. Dashed lines denote the closed form expression~(\ref{sintegral}) for the leading term.}
   \label{fig:fig2}
\end{figure}

In this work, we will consider circuits built out of the following elementary cycle of layers
\begin{equation}\label{constt}
U(t,h,\lambda)=U_X(i\lambda)U_X(h)U_{ZZ}(t)\,,
\end{equation}
for $t,h,\lambda$ real parameters. We can use this cycle to build more complicated rounds described by $\mathcal{U}$:
\begin{equation}\label{U}
    \mathcal{U}=U(t_p,h_p,\lambda_p)...U(t_1,h_1,\lambda_1)\,,
\end{equation}
with $t_r,h_r,\lambda_r,r\in\{1,...,p\}$ fixed real parameters. We will study the entanglement properties of a sub-system of large size $\ell$ after $n\to\infty$ identical rounds $\mathcal{U}$, in the thermodynamic limit $L\to\infty$. The order of limits is $\ell \ll n\ll L$. We define thus
\begin{equation}
    |\psi_n\rangle=\frac{1}{\sqrt{\mathcal{N}}}\mathcal{U}^n|\psi_0\rangle\,,
\end{equation}
with $\mathcal{N}$ a normalization such that $\langle\psi_n|\psi_n\rangle=1$. This kind of setup, for $p=1$, is illustrated in Fig.~\ref{fig:fig1}. Because each round is identical, our models have discrete time translation symmetry. While we use the particular structure of cycle and round defined in (\ref{constt}) and (\ref{U}), we note that our methods can be applied to any round built out of the layers in (\ref{layers}).

\textbf{\emph{Time evolution via M\"{o}bius transformations.}}---It was shown in Ref.\onlinecite{dreyer2021} that the action of the layers in (\ref{layers}) are particularly simple on coherent states. Coherent states are states of the form
\begin{equation}\label{coherent}
|\psi(\mathcal{A},f(k))\rangle=\mathcal{A}\prod_{k\in K^+_L}\left[1+f(k)c^\dagger(-k)c^\dagger(k)\right]|0\rangle\,,
\end{equation}
where $K_L=\frac{2\pi}{L}\left\{-\frac{L}{2}+\frac{1}{2},\cdots,\frac{L}{2}-\frac{1}{2}\right\}$ and $K^+_L\subset K_L$ contains all the positive momenta. Here, $\mathcal{A}$ normalizes the state and contains a phase, and $f(k)$ is an amplitude for fermions at momenta $k$ and $-k$. 

The crucial observation of Ref.~\onlinecite{dreyer2021} was that a coherent state stays a coherent state after the application of any of the layers in (\ref{layers}), but with modified $\mathcal{A}$ and $f(k)$. In particular, each of the layers in (\ref{layers}) transforms $f(k)$ by a M\"{o}bius transformation:
\begin{equation}
U_g(t)|\psi(\mathcal{A},f(k))\rangle=|\psi(\tilde{\mathcal{A}},\tilde{f}(k))\rangle\,,
\end{equation}
where $g=ZZ,YY,X$ and 
\begin{equation}\label{mob}
\tilde{f}(k)=F(f)=\frac{af(k)+b}{cf(k)+d}\,,
\end{equation}
where $a,b,c,$ and $d$ are complex functions of $t$ and $k$. We provide their explicit forms for the three kinds of layers in the supplemental material \cite{supp}. The transformation on $\mathcal{A}$ will not be needed in this work. Because the composition of M\"obius transformations is a M\"obius transformation, the action of $U(t,h,\lambda)$ and $\mathcal{U}$ can also be written as \eqref{mob}. It follows that $\mathcal{U}^n|\psi(\mathcal{A},f(k))\rangle$ also produces a coherent state for any $n$. Like any M\"obius transformation, the transformation associated with $\mathcal{U}$ can be packaged into a $2\times 2 $ matrix
\begin{equation}\label{Mobiusmatrix}
\mathcal{M}_k=\begin{pmatrix} a & b\\ c & d
\end{pmatrix}
\end{equation}
acting on the vector $\bigl( \begin{smallmatrix}f(k) \\ 1 \end{smallmatrix}\bigr)$: the new value $\tilde{f}(k)$ is given by the ratio of the two components of the resulting vector. Up to rescaling of the numerator and denominator in \eqref{mob} one can assume $\det(\mathcal{M}_k)=1$. The matrix of the M\"obius transformation associated with $\mathcal{U}^n$ is then simply obtained by repeated matrix multiplications $\mathcal{M}_k^n$. Therefore, in order to obtain the behaviour at large $n$ of $|\psi_n\rangle$, we need to study the fixed points of the M\"{o}bius transformation $\mathcal{M}_k$ associated with $\mathcal{U}$ and their stability.

Let us denote the state after $n$ rounds  by $|\psi(\mathcal{A}_n,f_n(k))\rangle$. The fixed points of the M\"{o}bius transformation are the two solutions to the quadratic equation 
\begin{equation}
f_{\infty}(k)=\frac{af_{\infty}(k)+b}{cf_{\infty}(k)+d}\,.
\end{equation}

We label these fixed points by $f_{\infty}^-(k)$ and $f_{\infty}^+(k)$ for each $k$. The stability of these fixed points are given by $|F'(f)|$: if $|F'(f)|_{f=f_{\infty}^{-}(k)}<1$, then $f_{\infty}^{-}(k)$ is a stable fixed point. Since a M\"{o}bius transformation can have at most only one stable fixed point, any choice of initial point $f_0(k)\neq f_\infty^+(k)$ will be attracted to $f_{\infty}^{-}(k)$ as $n\to\infty$, for that particular value of $k$. It can be shown that
\begin{equation}
|F'(f)|_{f=f_{\infty}^{-}}=\frac{1}{|F'(f)|_{f=f_{\infty}^{+}}}=\frac{|\mu_-(k)|}{|\mu_+(k)|}\,,
\end{equation}
where $\mu_-(k)$ and $\mu_+(k)$ with $|\mu_-(k)|\leq|\mu_+(k)|$ are the two eigenvalues of $\mathcal{M}_k$. So for $|\mu_-(k)|\neq |\mu_+(k)|$, there is a unique stable fixed point given by $f_{\infty}^{-}(k)$, while $f_{\infty}^{+}(k)$ is an unstable fixed point. On the contrary, if $|\mu_-(k)|= |\mu_+(k)|$, then there are no stable fixed points and $f_n(k)$ will not converge as $n\to\infty$ whenever $f_0(k)\neq f_\infty^\pm(k)$. 

We are going to show that these two alternatives completely determine the entanglement properties of the steady state. Let us call the values of $k$ for which $|\mu_-(k)|=|\mu_+(k)|$ ``critical". We will only need to consider $k\in[0,\pi]$ because $f(k)$ must be an antisymmetric function for (\ref{coherent}) to be consistent. If there are no critical $k$ in $[0,\pi]$, we will show that the steady state has area-law entanglement. For a 1D chain of $L$ sites, this means that the entanglement entropy $S_m(\ell)$ in the thermodynamic limit between the intervals $[1,\ell]$ and $[\ell+1,L]$, in the limit $L\to\infty$, saturates to a finite value when $\ell\to\infty$. If there is a finite number of critical $k$, then the steady state generically has log-law entanglement: $S_m(\ell)\sim\log(\ell)$. Finally, if there is a whole interval of critical $k$, then the steady state has volume-law entanglement: $S_m(\ell)\sim\ell$. 

In order to determine analytically whether or not such critical $k$ exist for a given M\"{o}bius transformation, we use the fact that $|\mu_-(k)|=|\mu_+(k)|$ if and only if the following two conditions are satisfied:
\begin{align}
\begin{split}\label{conditions}
&(i)\quad\mathfrak{I}\left(\mathrm{Tr}(\mathcal{M}_k)\right)=0\qquad(ii)\quad|\mathfrak{R}\left(\mathrm{Tr}(\mathcal{M}_k)\right)|\leq 2\,.
\end{split}
\end{align}
 We prove this in the supplemental material \cite{supp}. In the following, we will give two key examples of non-unitary circuits and show that, by studying their corresponding M\"{o}bius transformations, their phase diagrams can be determined exactly.

\textbf{\emph{Log-law to area-law transition.}}---We will now show that a log-law to area-law transition can occur in a simple circuit containing only one cycle (\ref{constt}), i.e. $p=1$ in \eqref{U}. For this circuit the M\"{o}bius transformation is given by
\begin{align}
\begin{split}
&\mathcal{M}_k=\frac{1}{\sqrt{1+|z_{k,t}|^2}}\begin{pmatrix} z_{k,t}e^{-2\lambda+2ih} & e^{-2\lambda+2ih} \\ -e^{2\lambda-2ih} & z_{k,t}^*e^{2\lambda-2ih}\end{pmatrix},
\end{split}
\end{align}
where we defined
\begin{equation}\label{zkt}
z_{k,t}=\frac{e^{2it}\tan(k/2)+\frac{e^{-2it}}{\tan(k/2)}}{2\sin(2t)}\,.
\end{equation}

To see whether a set of parameters $(t,h,\lambda)$ gives a steady state with log-law or area-law entanglement, we need to check for the existence of critical $k$ at which (\ref{conditions}) is satisfied. We will summarize the results here, and provide a more detailed derivation of the results in the supplemental material \cite{supp}. 

We find that if $\lambda=0$, both conditions in (\ref{conditions}) are satisfied for all $k$. Because there is a whole interval of critical $k$ (in fact, the entire range of $k$), the steady state exhibits volume-law entanglement, as expected in absence of measurements. For $\lambda>0$, the phase of the system depends on the condition $|\tan(2h)|>|\tan(2t)|$.
If this condition satisfied, then there are no critical points, and the system always satisfies area law. If it is not satisfied, then there is a critical value $\lambda_c$ such that for $0<\lambda<\lambda_c$ there is a unique critical $k$ given by $k=\arccos\left[\frac{\tan(2h)}{\tan(2t)}\right]$ for which both conditions of (\ref{conditions}) are satisfied, implying a log law for entanglement. For $\lambda>\lambda_c$ there are no critical $k$'s. We compute that the $t,h$-dependent value of $\lambda_c$ is given by
\begin{equation}
\lambda_c=\frac{1}{2}\mathrm{arcsinh}\left[\sqrt{\frac{2\sin^2(2t)}{\cos(4t)+\frac{\tan^2(2t)+\tan^2(2h)}{\tan^2(2t)-\tan^2(2h)}}}\right].
\end{equation}

Therefore, when $|\tan(2h)|<|\tan(2t)|$ holds, as we tune $\lambda$ through $\lambda_c$, the system demonstrates a log-law to area-law transition. Such transitions are well-known in free fermionic non-unitary circuits \cite{alberton2021,ippoliti2022fractal,chen2020,turkeshi2021}. 

Let us finally mention the case $h=t=\pi/4$. In this case we find for any $\lambda$ an interval $[k_c,\pi-k_c]$ of critical values of $k$ around $\pi/2$, with $0<k_c<\pi/2$ depending on $\lambda$. This yields a volume law phase for any $\lambda\geq 0$, but without any transition to an area-law behaviour, consistent with Ref.~\onlinecite{lu2021spacetime}. 

From the analysis above, we see that this kind of cycle cannot exhibit a volume-law to area-law transition at a finite $\lambda$. To find such a transition, we must consider a slightly more complex round.

\textbf{\emph{Volume-law to area-law transition}}---We now consider a round consisting of two of the cycles of the previous section:
\begin{equation}\label{twotimes}
\mathcal{U}=U(t_2,h_2,\lambda)U(t_1,h_1,\lambda)\,.
\end{equation}
For generic values of parameters, such circuits do not have volume-law to area-law MIPTs in $\lambda$. However, for the specific choice
\begin{equation}
t_1=h_1=\frac{\pi}{4}-x\qquad t_2=h_2=\frac{\pi}{4}+x,
\label{definition_of_x}
\end{equation}
where $x\in[0,\pi/4]$, we will show that there exists an $x$-dependent $\lambda_c$ such that for $\lambda<\lambda_c$, the steady state demonstrates volume-law entanglement, while for $\lambda>\lambda_c$, the steady-state demonstrates area-law entanglement.

From the definition of $z_{k,t}$ in (\ref{zkt}), we find that $z_{k,t_1}=-z_{k,t_2}^*$. Therefore, $\mathcal{M}_k$ for (\ref{twotimes}) simplifies to
\begin{align}
\begin{split}
&\mathcal{M}_k=\frac{1}{1+|z_{k,t_1}|^2}\\
&\times\begin{pmatrix} |z_{k,t_1}|^2e^{-4\lambda}-e^{4ix} & z_{k,t_1}^*(e^{-4\lambda}+e^{4ix}) \\ -z_{k,t_1}(e^{4\lambda}+e^{-4ix}) & |z_{k,t_1}|^2e^{4\lambda}-e^{-4ix}\end{pmatrix}.
\end{split}
\end{align}
To determine the critical $k$, we compute
\begin{equation}
\mathrm{Tr}(\mathcal{M}_k)=2\frac{|z_{k,t_1}|^2\cosh(4\lambda)-\cos(4x)}{1+|z_{k,t_1}|^2}.
\end{equation}
This quantity is always real, so condition $(i)$ of (\ref{conditions}) is satisfied for all $k$. Condition $(ii)$ can be written as
\begin{equation}\label{condvolume}
\tan^2(k/2)+\frac{1}{\tan^2(k/2)}\leq \frac{4\cos^4(2x)}{\sinh^2(2\lambda)}+2\cos(4x).
\end{equation}
For this to be true for at least one value of $k$, we need it to be true for $k=\frac{\pi}{2}$, which minimizes the left-hand side. Plugging in $k=\frac{\pi}{2}$, we find that the interval of critical $k$ disappears when $\lambda>\lambda_c$ where
\begin{equation}\label{lambdacvolume}
\lambda_c=\frac{1}{2}\mathrm{arcsinh}\left[\frac{\cos^2(2x)}{\sin(2x)}\right].
\end{equation}

Therefore, the steady state demonstrates area-law entanglement for $\lambda>\lambda_c$. For $\lambda<\lambda_c$, (\ref{condvolume}) is satisfied for $k\in[k_c,\pi-k_c]$, with $k_c$ obeying (\ref{condvolume}) with equality. Within this interval, $f_n(k)$ does not converge to a stable fixed point, and we will now show that this leads to volume-law entanglement.

\textbf{\emph{Behavior of entanglement entropy}}---We begin by defining the kinds of entanglement entropy we will compute. Given a state $|\psi\rangle$ on a spin chain of size $L$, the reduced density matrix $\rho_{\ell}$ on $[1,\ell]$ is given by
\begin{equation}
\rho_{\ell}=\mathrm{Tr}_{\ell+1,\dots,L}\left(|\psi\rangle\langle\psi|\right).
\end{equation}
We define the entanglement entropies $S_m(\ell)$ as
\begin{equation}
\begin{aligned}
     S_0(\ell)&=\log\, {\rm rank}[\rho]\,,\qquad \text{Hartley}\\
     S_1(\ell)&=-\mathrm{Tr}[\rho \log \rho]\,,\qquad \text{von Neumann}\\
     S_m(\ell)&=\frac{\log \mathrm{Tr}[\rho^m]}{1-m}\,,\qquad m\geq 2\,,\qquad \text{R\'enyi}\,.
\end{aligned}
\end{equation}

While it is difficult to obtain the exact behavior of $S_m(\ell)$ at intermediate times, we can compute the coefficient of the volume-law ($\mathcal{O}(\ell)$) contribution to $S_m(\ell)$ in the $n\to\infty$ limit. To that end, we introduce the correlation matrix $\Gamma$ on the subsystem $[1,2,\dots \ell]$, given by
\begin{equation}
\langle\psi|\begin{pmatrix} a_{2j-1}\\ a_{2j}\end{pmatrix}\cdot\begin{pmatrix} a_{2k-1} & a_{2k}\end{pmatrix}|\psi\rangle=\delta_{j,k}+i\Gamma_{jk},
\end{equation}
with the Majorana fermions $a_{2j-1}=c_j+c_j^\dagger$ and  $a_{2j}=i(c_j-c_j^\dagger)$. $\Gamma$ has a block-Toeplitz structure:
\begin{equation}
    \Gamma=\left(\begin{matrix}
        \Pi_0 & \Pi_1&\hdots & \Pi_{\ell-1}\\ \Pi_{-1}&\Pi_0&\hdots \\
        \vdots & & \ddots& \vdots\\
        \Pi_{1-\ell}&\hdots &\hdots &\Pi_0
    \end{matrix} \right)\,\qquad  \Pi_j=\left(\begin{matrix}
        -\varphi_j & \psi_j\\- \psi_{-j}& \varphi_j
    \end{matrix} \right)\,,
    \end{equation}
where from \eqref{coherent}, $\varphi_j$ and $\psi_j$ are computed to be
\begin{align}
\begin{split}\label{coefficients}
    \varphi_j&=\frac{i}{2\pi}\int_{-\pi}^\pi dk e^{-ikj}\frac{f_n(k)+f_n(k)^*}{1+|f_n(k)|^2}\\
    \psi_j&=\frac{1}{2\pi}\int_{-\pi}^\pi dk e^{-ikj} \frac{f_n(k)-f_n(k)^*+|f_n(k)|^2-1}{1+|f_n(k)|^2}\,.
\end{split}
\end{align}
We find that $\varphi_j$ and $\psi_j$ converge to some $\bar{\varphi}_j$ and $\bar{\psi}_j$ respectively as $n\to\infty$. It is convenient to define $\hat{\varphi}_j$ and $\hat{\psi}_j$ through
\begin{align}
\begin{split}
\bar{\varphi}_j=\frac{1}{2\pi}\int_{-\pi}^{\pi}dk e^{-ikj}\hat{\varphi}(k)\,\quad \bar{\psi}_{j}&=\frac{1}{2\pi}\int_{-\pi}^{\pi}dk e^{-ikj}\hat{\psi}(k).
\end{split}
\end{align}

When $f_n(k)$ converges to a stable fixed point, we can replace $f_n(k)$ in (\ref{coefficients}) by $f_{\infty}^-(k)$. Notice that if there is a critical value of $k$, then $f_{\infty}^-(k)$ generically fails to be smooth because it jumps between the two fixed points of the M\"{o}bius transformation. This leads to $\varphi_j$ and $\psi_j$, which are Fourier transformations of functions of $f_{\infty}^-(k)$, to have power-law correlations. Power-law correlations of the Majorana fermions in real-space implies log-law entanglement. On the other hand, if there are no critical $k$, $f_{\infty}^-(k)$ is smooth and real-space correlations decay exponentially, implying area-law entanglement \cite{hastings2007,brandao2015}.

If $k\in[k_c,\pi-k_c]$ is critical, then $f_n(k)$ in this momentum range does not converge and depends on both the initial state $f_0(k)$ and the cycle number $n$. Let us choose the initial state $f_n(k)=0$ for all $k$ which, in the spin language, corresponds to the state with all spins in the $+1$ eigenstate of $\sigma^x_j$. For this initial state, we have $f_n(k)=x_n(k)/y_n(k)$ with
\begin{equation}
\begin{pmatrix} x_n(k)\\ y_n(k)\end{pmatrix}=\mathcal{M}_k^n\cdot\begin{pmatrix} 0\\ 1\end{pmatrix}\,.
\end{equation}
%
Since $\mathrm{det}(\mathcal{M}_k)=1$, if $|\mu_-(k)|=|\mu_+(k)|$, then $\mu_{\pm}(k)=e^{\pm i\theta_k}$ for $k\in[k_c,\pi-k_c]$, where
\begin{equation}
\theta_k=\arccos\left[\frac{|z_{k,t_1}|^2\cosh(4\lambda)-\cos(4x)}{1+|z_{k,t_1}|^2}\right]\,.
\end{equation}
Diagonalizing $\mathcal{M}_k$, we find
\begin{equation}\label{fnk}
f_n(k)=\frac{b(k)\sin(n\theta_k)}{(-a(k)+\cos\theta_k)\sin(n\theta_k)+\sin(\theta_k)\cos(n\theta_k)},
\end{equation}
where $a$ and $b$ are matrix elements of $\mathcal{M}_k$ as defined in (\ref{Mobiusmatrix}), and we have indicated their $k$ dependence explicitly. We see that in this case, $f_n(k)$ does not converge to a stable fixed point as $n\to\infty$ and instead keeps oscillating. To compute $\hat{\varphi}(k)$ and $\hat{\psi}(k)$, we separate the slowly varying and quickly oscillating parts of $f_n(k)$, and define
\begin{equation}\label{mathfrak}
    \mathfrak{f}(k,u)=\frac{b(k) \sin u}{(-a(k)+\cos\theta_k)\sin u+\sin\theta_k \cos u}\,.
\end{equation}
As shown in the Supplemental Material \cite{supp} we compute $\hat{\varphi}(k)$ and $\hat{\psi}(k)$ by averaging over the fast oscillations:
\begin{align}
\begin{split}\label{volume}
    \hat{\varphi}(k)&=\frac{i}{2\pi}\int_0^{2\pi}du \frac{\mathfrak{f}(k,u)+\mathfrak{f}(k,u)^*}{1+|\mathfrak{f}(k,u)|^2}\\
    \hat{\psi}(k)&=\frac{1}{2\pi}\int_0^{2\pi}du \frac{\mathfrak{f}(k,u)-\mathfrak{f}(k,u)^*+|\mathfrak{f}(k,u)|^2-1}{1+|\mathfrak{f}(k,u)|^2}\,.
\end{split}
\end{align}
We now follow Ref.\onlinecite{calabrese2005} to compute the entanglement entropy. Repeating the calculations therein, we find the following leading behavior when $\ell\to\infty$:
\begin{align}
\begin{split}\label{sintegral}
S_m(\ell)&=\frac{\ell}{2\pi}\int_{-\pi}^{\pi}dk H_m\left(\sqrt{|\hat{\varphi}(k)|^2+|\hat{\psi}(k)|^2}\right)\\
&+\mathcal{O}(\log\ell),
\end{split}
\end{align}
with the function $H_m(x)$ given by
\begin{align}
\begin{split}
H_0(x)&=\mathbbm{1}_{x\in(-1,1)}\\
H_1(x)&=-\frac{1+x}{2}\log\frac{1+x}{2}-\frac{1-x}{2}\log\frac{1-x}{2}\\
H_m(x)&=\frac{1}{1-m}\log\left[\left(\frac{1+x}{2}\right)^m+\left(\frac{1-x}{2}\right)^m\right],
\end{split}
\end{align}
for $m\geq 2$. When $f_n(k)$ converges to a stable fixed point, we plug $f_{\infty}^-(k)$ into the definition of $\hat{\varphi}(k)$ and $\hat{\psi}(k)$, and obtain 
\begin{equation}
|\hat{\varphi}(k)|^2+|\hat{\psi}(k)|^2=1.
\end{equation}
Since $H_m(1)=0$, these values of $k$ do not contribute to the coefficient of the $\mathcal{O}(\ell)$ term in the entanglement entropy. On the other hand, when $f_n(k)$ does not converge to a stable fixed point, we can show from (\ref{mathfrak}) that \cite{supp}
\begin{equation}\label{ineq}
|\hat{\varphi}(k)|^2+|\hat{\psi}(k)|^2<1.
\end{equation}
Therefore, these momenta do contribute to the $\mathcal{O}(\ell)$ term in the entanglement entropy. In summary, we obtain
\begin{equation}
    S_m(\ell)=\begin{cases}
        s_m(\lambda)\ell\,,\qquad \text{if }\lambda<\lambda_c\\
        \mathcal{O}(\ell^0)\,,\qquad \text{if }\lambda>\lambda_c\,,
    \end{cases}
\end{equation}
with $s_m(\lambda)$ a coefficient that we can compute through the integral in (\ref{sintegral}) (see supplemental material \cite{supp}). We compare these exact calculations with numerical computations of entanglement entropy in Fig.~\ref{fig:fig2}. In the limit $\lambda\to\lambda_c$, we define the critical exponent $\nu$ by the leading behaviour of $s_m(\lambda)$ when $\lambda\to\lambda_c$:
\begin{equation}
    s_m(\lambda)=a_m(\lambda_c-\lambda)^\nu+\mathcal{O}((\lambda_c-\lambda)^{\nu'}),
\end{equation}
where $\nu'>\nu$. We obtain analytically
\begin{align}
\begin{split}
s_0(\lambda_c-\lambda_c)&\sim\left(\lambda_c-\lambda\right)^{1/2}\\
s_1(\lambda_c-\lambda)&\sim\left(\lambda_c-\lambda\right)\log\left(\lambda_c-\lambda\right)\\
s_m(\lambda_c-\lambda)&\sim\left(\lambda_c-\lambda\right)\,,
\end{split}
\end{align}
with coefficients given in the supplemental material \cite{supp}. Therefore, $\nu=\frac{1}{2}$ for $S_0(\ell)$ and $\nu=1$ for $S_m(\ell)$, where $m\geq 1$, with marginal logarithmic corrections at $m=1$ \cite{affleck1989,eggert1996}. This differs from the exponents in the literature for the Haar random case with projective measurements, found to be $4/3$ for $S_0$ and $\approx 1.3$ for $S_1$ \cite{skinner2019,li2019,zabalo2020,choi2020,gullans2020scalable,block2022measurement,lunt2020measurement,lu2021spacetime}. We also compute subleading terms $\sim(\lambda_c-\lambda)^{3/2}$ that would lead to overestimating the exponent $\nu=1$ when using data for small $\lambda_c-\lambda$. By computing the entanglement entropy at $\lambda=\lambda_c$ numerically, we observe that the $\mathcal{O}(\log\ell)$ term vanishes at the critical point, indicating that the central charge is $c=0$, consistent with Ref.~\onlinecite{turkeshi2021,alberton2021}.
   
\textbf{\emph{Discussion}}---We presented a general framework for obtaining exact results on steady-states of clean Gaussian non-unitary circuits with discrete time translation symmetry using fixed points of M\"{o}bius transformations. A few comments are in order. First, it was shown in Ref.~\onlinecite{gullans2020,fidkowski2021} that entanglement transitions can also be interpreted as purification transitions, if the initial state is a mixed state rather than a pure state. This interpretation can also be seen in our analysis: when a M\"{o}bius transformation has a single stable fixed point for all $k$, the steady state is independent of the initial state. Therefore, a density matrix that is a linear combination of different coherent state density matrices will be mapped, as $n\to\infty$, to the density matrix of the pure state given by $f_{\infty}^-(k)$. On the other hand, when the M\"{o}bius transformation does not have any fixed points for $k\in[k_1,k_2]$, $f_n(k)$ in this interval continues to depend on the initial state at large $n$. Therefore, even as $n\to\infty$, a mixed state density matrix will remain mixed, and does not purify.

Second, we note that the distinction governed by conditions \eqref{conditions} is equivalent to a statement on the reality of the single-particle energies of the effective Hamiltonian $H$ defined by
\begin{equation}
    \mathcal{U}=e^{iH}\,.
\end{equation}
Indeed, $H$ is block-diagonal in the space generated by Fourier modes $k,-k$, and although its eigenvalues in that subspace are not directly related to those of $\mathcal{M}_k$, they are real if and only if $k$ is critical. Therefore, increasing the non-unitarity of the model through $\lambda$ has a similar effect as in the continuous-time model of Ref. \onlinecite{legal2022}.

There are many future directions related to this work. While we only used M\"{o}bius transformations to study two simple kinds of cycles, the general framework can be used to study a broad range of dynamics. It would be interesting to explore other kinds of circuits built out of the layers in (\ref{layers}), with both real and imaginary time. In addition, while we have focused on steady states in this work, it would be interesting to study the intermediate-time dynamics of these systems. It would also be interesting to further pinpoint the exact nature of the critical points. In forthcoming work, we study the effects of breaking discrete time translation symmetry with temporal disorder. The circuits maintain translation symmetry, and the steady state is obtained from an infinite number of M\"{o}bius transformations with random noise. Another way to break discrete time translation symmetry is by removing postselection, which leads to randomness in the sign of $\lambda$. For experimental applications, this is an important direction for future work.

C.Z. thanks David Huse, Tim Hsieh, and Soonwon Choi for helpful discussions. E.G. thanks Adam Nahum and Lorenzo Piroli for interesting discussions. C.Z. acknowledges support from the University of Chicago Bloomenthal Fellowship and the National
Science Foundation Graduate Research Fellowship under Grant No. 1746045. E.G. acknowledges support from the Kadanoff Center for Theoretical Physics at University of Chicago, and from the Simons Collaboration on Ultra-Quantum Matter.
\bibliography{bibliography}

\end{document}


\title{Supplemental Material}
\author{Etienne Granet and Carolyn Zhang}
\affiliation{Department of Physics, Kadanoff Center for Theoretical Physics, University of Chicago, Chicago, Illinois 60637,  USA}
\author{Henrik Dreyer}
\affiliation{Quantinuum, Leopoldstrasse 180, 80804 Munich, Germany}

\maketitle
\section{Realization of the weak measurement}
The non-unitary operator $e^{\lambda\sigma^x_j}$ can be realized (up to an unimportant normalization factor) with an ancilla spin and a post-selection. Namely we have, with $\sigma^x_j$ acting on the spin in the spin chain and $\sigma^y_a$ acting on the ancilla,
\begin{equation}
    e^{\lambda\sigma^x_j}\propto {}_a\langle 0|e^{i\left(\frac{\theta+\theta'}{2}+\frac{\theta-\theta'}{2}\sigma^x_j \right)\sigma^y_a} |0\rangle_a\,,\qquad \text{with }e^{2\lambda}=\frac{\cos \theta}{\cos\theta'} \,.
\end{equation}
Notice that we post-select the ancilla to be in the $+1$ eigenstate of $\sigma^y_a$. The measurement outcome of the ancilla determines the sign of $\lambda$, so we must post-select after every cycle to maintain the same sign.

\section{M\"{o}bius transformations}
The M\"{o}bius transformations for the layers in Eq. (1) of the main text were derived in Ref.~\onlinecite{dreyer2021}. We list them here for convenience:
\begin{align}
\begin{split}
U_{ZZ}(t): \tilde{f}(k)&=\frac{[1+\tan^2(k/2)e^{4it}]f(k)+i\tan(k/2)(1-e^{4it})}{-i\tan(k/2)(1-e^{4it})f(k)+\tan^2(k/2)+e^{4it}}\\
U_{YY}(t):\tilde{f}(k)&=\frac{[1+\tan^2(k/2)e^{4it}]f(k)-i\tan(k/2)(1-e^{4it})}{i\tan(k/2)(1-e^{4it})f(k)+\tan^2(k/2)+e^{4it}}\\
U_X(t):\tilde{f}(k)&=e^{4it}f(k)\,.
\end{split}
\end{align}

\section{Proof of Eq. (12)}
We now prove that the two eigenvalues $\mu_-(k)$ and $\mu_+(k)$ of $\mathcal{M}_k$ with $\det \mathcal{M}_k=1$ have equal magnitude if and only if 
\begin{align}
\begin{split}\label{conditionssupp}
&(i)\qquad\mathfrak{I}\left(\mathrm{Tr}(\mathcal{M}_k)\right)=0\\
&(ii)\qquad|\mathfrak{R}\left(\mathrm{Tr}(\mathcal{M}_k)\right)|\leq 2\,.
\end{split}
\end{align}
Let $K=\frac{\mu_-(k)}{\mu_+(k)}$. We use the fact that
\begin{equation}
\sqrt{K}+\frac{1}{\sqrt{K}}=\frac{\mathrm{Tr}(\mathcal{M}_k)}{\sqrt{\mathrm{det}(\mathcal{M}_k)}}=\mathrm{Tr}(\mathcal{M}_k)\,.
\end{equation}
Now setting $K=\kappa e^{i\alpha}$ with $\kappa>0$, the two eigenvalues have equal magnitude if and only if $\kappa=1$, and we have
\begin{equation}
\mathfrak{I}\left(\mathrm{Tr}(\mathcal{M}_k)\right)=\left(\sqrt{\kappa}-\frac{1}{\sqrt{\kappa}}\right)\sin\frac{\alpha}{2}\,.
\end{equation}
The above quantity is zero if and only if $\kappa=1$ or $\sin(\alpha/2)=0$. Similarly, we have
\begin{equation}
\mathfrak{R}\left(\mathrm{Tr}(\mathcal{M}_k)\right)=\left(\sqrt{\kappa}+\frac{1}{\sqrt{\kappa}}\right)\cos\frac{\alpha}{2}\,.
\end{equation}
If $\kappa=1$, then the right-hand side is smaller than or equal to $2$ in absolute value. This proves that $\kappa=1$ implies \eqref{conditionssupp}. If now $\kappa\neq 1$ but $\sin(\alpha/2)=0$, then $\cos(\alpha/2)=\pm 1$ and the right-hand side is strictly larger than $2$ in absolute value. Hence if $\kappa\neq 1$, at least one of the two conditions in \eqref{conditionssupp} is not satisfied, which proves the equivalence.

\section{Log-law to area-law transition}
We consider here the setup given by the matrix in Eq. (12). Defining
\begin{equation}
    z_{k,t}=\rho e^{i\phi}\,,\qquad \rho>0\,,
\end{equation}
we have
\begin{equation}\label{cond22}
\begin{aligned}
    \Re \tr \mathcal{M}_k&=\frac{2\rho\cos(\phi+2h)\cosh(2\lambda)}{\sqrt{1+\rho^2}}\\
 \Im \tr \mathcal{M}_k&=-\frac{2\rho\sin(\phi+2h)\sinh(2\lambda)}{\sqrt{1+\rho^2}}\,.
\end{aligned}
\end{equation}
From the definition of $z_{k,t}$, we have
\begin{equation}
    \rho^2=\frac{\tan^2(k/2)+\tan^{-2}(k/2)+2\cos(4t)}{4\sin^2(2t)}\,,
\end{equation}
and
\begin{equation}\label{phi}
    \tan \phi=-\cos k \tan(2t)\,.
\end{equation}
The condition $\sin(\phi+2h)=0$ is equivalent to $\tan\phi=-\tan(2h)$, which thus gives the equation
\begin{equation}
    \cos k=\frac{\tan(2h)}{\tan(2t)}\,.
\end{equation}
For this to have a solution one needs $|\tan(2h)|\leq |\tan(2t)|$. Hence for $\lambda\neq 0$, the condition (i) of \eqref{conditionssupp} is only satisfied for $k_c=\pm\arccos \frac{\tan(2h)}{\tan(2t)}$ when $|\tan(2h)|\leq |\tan(2t)|$. Then the condition (ii) of \eqref{conditionssupp} yields at $k=k_c$
\begin{equation}
    \frac{\rho^2}{1+\rho^2}\cosh^2(2\lambda)\leq 1\,.
\end{equation}
Using the expression for $\rho^2$ and
\begin{equation}
    \tan^2(k_c/2)=\frac{\tan(2t)-\tan(2h)}{\tan(2t)+\tan(2h)}\,,
\end{equation}
one finds then that it is equivalent to $\lambda\leq \lambda_c$ with $\lambda_c$ given by Eq. (14) in the main text.


\section{Proof of Eq. (31)}\label{volumeproof}
In this appendix, we prove Eq. (31) in the main text, where we obtain $\hat{\phi}(k)$ and $\hat{\psi}(k)$ by averaging over $u$:
\begin{align}
\begin{split}\label{volumesupp}
    \hat{\varphi}(k)&=\frac{i}{2\pi}\int_0^{2\pi}du \frac{\mathfrak{f}(k,u)+\mathfrak{f}(k,u)^*}{1+|\mathfrak{f}(k,u)|^2}\\
    \hat{\psi}(k)&=\frac{1}{2\pi}\int_0^{2\pi}du \frac{\mathfrak{f}(k,u)-\mathfrak{f}(k,u)^*+|\mathfrak{f}(k,u)|^2-1}{1+|\mathfrak{f}(k,u)|^2}\,.
\end{split}
\end{align}
For a smooth bounded function $F(k,u)$ that is $2\pi$ periodic in $u$, we can decompose for $\epsilon>0$
\begin{equation}
\begin{aligned}
    \int_{-\pi}^\pi F(k,n\theta_k)\D{k}&=\sum_{m=0}^{2\pi/\epsilon} \int_{-\pi+m\epsilon}^{-\pi+(m+1)\epsilon}\D{k} F(k,n\theta_k)\\
    &=\sum_{m=0}^{2\pi/\epsilon} \int_{-\pi+m\epsilon}^{-\pi+(m+1)\epsilon}\D{k} F(-\pi+m\epsilon,n\theta_k)+\mathcal{O}(\epsilon)\\
    &=\sum_{m=0}^{2\pi/\epsilon} \frac{1}{n\theta'_{-\pi+m\epsilon}}\int_{n\theta_{-\pi+m\epsilon}}^{n\theta_{-\pi+(m+1)\epsilon}}\D{u} F(-\pi+m\epsilon,u)+\mathcal{O}(\epsilon)\,.
\end{aligned}
\end{equation}
Importantly, the $\mathcal{O}(\epsilon)$ can be bounded independently of $n$ from the assumptions on $F$. Each of these terms in the sum has a limit when $n\to\infty$ given by
\begin{equation}
    \underset{n\to\infty}{\lim}\, \frac{1}{n\theta'_{-\pi+m\epsilon}}\int_{n\theta_{-\pi+m\epsilon}}^{n\theta_{-\pi+(m+1)\epsilon}}\D{u} F(-\pi+m\epsilon,u)=\frac{\theta_{-\pi+(m+1)\epsilon}-\theta_{-\pi+m\epsilon}}{2\pi\theta'_{\pi+m\epsilon}}\int_{0}^{2\pi}\D{u} F(-\pi+m\epsilon,u)\,.
\end{equation}
Hence we obtain 
\begin{equation}
    \begin{aligned}
        \int_{-\pi}^\pi F(k,n\theta_k)\D{k}&=\sum_{m=0}^{2\pi/\epsilon} \frac{\theta_{-\pi+(m+1)\epsilon}-\theta_{-\pi+m\epsilon}}{2\pi\theta'_{\pi+m\epsilon}}\int_{0}^{2\pi}\D{u} F(-\pi+m\epsilon,u)+\mathcal{O}(\epsilon)+\mathcal{O}(1/n)\\
         &=\frac{\epsilon}{2\pi}\sum_{m=0}^{2\pi/\epsilon} \int_{0}^{2\pi}\D{u} F(-\pi+m\epsilon,u)+\mathcal{O}(\epsilon)+\mathcal{O}(1/n)\\
         &=\frac{1}{2\pi}\int_{-\pi}^\pi \D{k} \int_{0}^{2\pi}\D{u} F(k,u)+\mathcal{O}(\epsilon)+\mathcal{O}(1/n)\,.
    \end{aligned}
\end{equation}
We can now apply this to $\varphi_j$ and $\psi_j$ with $f_n(k)$ given in Eq. (29) of the main text:
\begin{equation}\label{fnksupp}
f_n(k)=\frac{b(k)\sin(n\theta_k)}{(-a(k)+\cos\theta_k)\sin(n\theta_k)+\sin(\theta_k)\cos(n\theta_k)},
\end{equation}
For example for $\varphi_j$ we would choose
\begin{equation}
    F(k,u)=i e^{-ikj}\frac{\mathfrak{f}(k,u)+\mathfrak{f}(k,u)^*}{1+|\mathfrak{f}(k,u)|^2}\,,
\end{equation}
with the definition 
\begin{equation}\label{mathfrak}
    \mathfrak{f}(k,u)=\frac{b(k) \sin u}{(-a(k)+\cos\theta_k)\sin u+\sin\theta_k \cos u}\,,
\end{equation}
and we see that $F(k,u)$ is indeed smooth and bounded in absolute value by $1$. Hence we obtain that $\varphi_j$ and $\psi_j$ converge when $n\to\infty$ to $\bar{\varphi}_j$ and $\bar{\psi}_j$ with $\hat{\varphi}(k)$ and $\hat{\psi}(k)$ given by \eqref{volumesupp}.

\section{Proof of Eq. (32)}\label{EEproof}
We follow Ref.\onlinecite{calabrese2005} to prove Eq. (32) in the main text. Denoting $\pm i\nu_p$ the eigenvalues of $\Gamma$, the entanglement entropies $S_m(\ell)$ are
\begin{equation}
    S_m(\ell)=\sum_{p=1}^\ell H_m(\nu_p)\,.
\end{equation}
Defining
\begin{equation}
    D(\lambda)=\det[i\lambda\, {\rm Id}_{2\ell}-\Gamma]\,,
\end{equation}
we have $D(\lambda)=\prod_{p=1}^\ell (\lambda^2-\nu_p^2)$, and so $S_m(\ell)$ can be written as a contour integral in the upper-half plane of $\partial_\lambda \log D(\lambda)$ multiplied by $H_m(\lambda)$, since each $\nu_p$ is a simple pole of $\partial_\lambda \log D(\lambda)$. Let us introduce the $2\times 2$ matrices $\hat{\Pi}(k)$ by
\begin{equation}
    \bar{\Pi}_j=\frac{1}{2\pi}\int_{-\pi}^{\pi}dk e^{-ikj}\hat{\Pi}(k)\,,
\end{equation}
where $\bar{\Pi}_j$ denotes the limit $n\to\infty$ of $\Pi_j$. It reads
\begin{equation}
    \hat{\Pi}(k)=\left(\begin{matrix}
    -\hat{\varphi}(k) & \hat{\psi}(k)\\ -\hat{\psi}(-k)^* & \hat{\varphi}(k)
    \end{matrix}\right)\,.
\end{equation}
Now, using Szego's lemma, we have when $\ell\to\infty$
\begin{equation}
    \log D(\lambda)=\frac{\ell}{2\pi}\int_0^{2\pi} \D{k} \log \det[i\lambda\, {\rm Id}_2-\hat{\Pi}(k)]+\mathcal{O}(\log \ell)\,.
\end{equation}
We find, using $\hat{\psi}(-k)=\hat{\psi}(k)^*$
\begin{equation}
     \det[i\lambda\, {\rm Id}_2-\hat{\Pi}(k)]=|\hat{\psi}(k)|^2+|\hat{\varphi}(k)|^2-\lambda^2\,.
\end{equation}
Hence $\partial_\lambda \log D(\lambda)$ has simple poles at
\begin{equation}
    \lambda=\pm \sqrt{|\hat{\psi}(k)|^2+|\hat{\varphi}(k)|^2}\,.
\end{equation}
Hence, carrying the contour integral yields Eq. (32) of the main text.

\section{Proof of Eq. (35)}\label{inequalityproof}
In this appendix, we prove the statement in the main text that
\begin{equation}\label{inequality}
|\hat{\varphi}(k)|^2+|\hat{\psi}(k)|^2<1
\end{equation}
when $k$ is critical. Since $x\to x^2$ is a strictly convex function, we have for any function $F(u)$
\begin{equation}
    \left(\frac{1}{2\pi}\int_0^{2\pi}\D{u}F(u) \right)^2\leq \frac{1}{2\pi}\int_0^{2\pi} \D{u} F(u)^2\,,
\end{equation}
with equality if and only if $F(u)$ is constant. We then write
\begin{equation}
\begin{aligned}
    |\hat{\varphi}(k)|^2+|\hat{\psi}(k)|^2=&\left(\frac{1}{2\pi}\int_0^{2\pi}\D{u}\frac{2\Re\mathfrak{f}(k,u)}{1+|\mathfrak{f}(k,u)|^2} \right)^2+\left(\frac{1}{2\pi}\int_0^{2\pi}\D{u}\frac{|\mathfrak{f}(k,u)|^2-1}{1+|\mathfrak{f}(k,u)|^2} \right)^2\\
    &+\left(\frac{1}{2\pi}\int_0^{2\pi}\D{u}\frac{2\Im \mathfrak{f}(k,u)}{1+|\mathfrak{f}(k,u)|^2} \right)^2\,.
\end{aligned}
\end{equation}
Since $\mathfrak{f}(k,u)$ is not constant in $u$, we have
\begin{equation}
    |\hat{\varphi}(k)|^2+|\hat{\psi}(k)|^2<\frac{1}{2\pi}\int_0^{2\pi}\D{u}\left[ \left(\frac{2\Re\mathfrak{f}(k,u)}{1+|\mathfrak{f}(k,u)|^2} \right)^2+\left(\frac{2\Im\mathfrak{f}(k,u)}{1+|\mathfrak{f}(k,u)|^2} \right)^2+\left(\frac{|\mathfrak{f}(k,u)|^2-1}{1+|\mathfrak{f}(k,u)|^2} \right)^2\right]\,.
\end{equation}
The integrand simplifies to $1$, yielding \eqref{inequality}.

\section{Critical exponents}
The entanglement entropy is given by
\begin{equation}\label{ent}
    S_m(\ell)=\frac{\ell}{2\pi}\int_{-\pi}^{\pi}\D{k} H_m\left(\sqrt{|\hat{\varphi}(k)|^2+|\hat{\psi}(k)|^2} \right)+\mathcal{O}(\log \ell)\,.
\end{equation}
Since $|\hat{\varphi}(k)|^2+|\hat{\psi}(k)|^2=1$ when $k$ is not critical, and since $H_m(1)=0$, only the critical values of $k$ contribute to the volume law scaling. Hence we have 
\begin{equation}\label{ent}
    S_m(\ell)=\frac{\ell}{\pi}\int_{k_c}^{\pi-k_c}\D{k} H_m\left(\sqrt{|\hat{\varphi}(k)|^2+|\hat{\psi}(k)|^2} \right)+\mathcal{O}(\log \ell)\,.
\end{equation}
where $k_c$ satisfies
\begin{equation}\label{kstar}
\tan(k_c/2)^2+\tan(k_c/2)^{-2}=\frac{4\cos^4(2x)}{\sinh^2(2\lambda)}+2\cos (4x)\,.
\end{equation}
Recall that the entanglement entropy $S_m(\ell)$ has the following behaviour at large $\ell$
\begin{equation}
    S_m(\ell)=\begin{cases}
        s_m(\lambda)\ell\,,\qquad \text{if }\lambda<\lambda_c\\
        \mathcal{O}(\ell^0)\,,\qquad \text{if }\lambda>\lambda_c\,,
    \end{cases}
\end{equation}
with $s_m(\lambda)$ some coefficient. We define the critical exponent $\nu$ by the leading behaviour of $s_m(\lambda)$ when $\lambda\to\lambda_c$
\begin{equation}
    s_m(\lambda)=a_m(\lambda_c-\lambda)^\nu+\mathcal{O}((\lambda_c-\lambda)^{\nu'})\,,
\end{equation}
where $\nu'>\nu$. This critical behaviour can be studied performing an expansion of \eqref{ent} when $\lambda\to\lambda_c$. To that end, we need the dominant expression of $k_c$ and $\theta_k$ when $\lambda\to\lambda_c$. Firstly, from \eqref{kstar} we find
\begin{equation}\label{kcexpand}
    k_c=\frac{\pi}{2}-\frac{2\sin(2x)}{\sqrt{\tanh(2\lambda_c)}}\sqrt{\lambda_c-\lambda}+\mathcal{O}(\lambda_c-\lambda)\,.
\end{equation}
We now perform an expansion of $\theta_k$ for $k\in[k_c,\pi-k_c]$. We note that as $\lambda\to\lambda_c$, we have $k_c\to\pi/2$, so the values of $k\in[k_c,\pi-k_c]$ also come close to $\pi/2$. This requires thus an expansion of $\theta_k$ in $(k-\pi/2)^2$ and $\lambda_c-\lambda$. We find
\begin{equation}\label{thetak}
    \theta_k=\sqrt{8\sin^2(2x)\sinh(4\lambda_c)(\lambda_c-\lambda)-2\cos^2(2x)[\cosh(4\lambda_c)+\cos(4x)](k-\tfrac{\pi}{2})^2}+\mathcal{O}(\lambda_c-\lambda)+\mathcal{O}(k-\tfrac{\pi}{2})^2\,.
\end{equation}
We then need the behaviour of $\hat{\varphi},\hat{\psi}$ perturbatively in $\lambda_c-\lambda$. For $a,b,c,d,e,f$ constants and $\theta\to 0$ we have the expansion 
\begin{equation}\label{exp}
   \frac{1}{2\pi} \int_0^{2\pi}\D{u} \frac{a\sin^2 u+b\theta \sin u \cos u+c\theta^2}{d\sin^2 u+e\theta \sin u\cos u+f \theta^2}=\frac{a}{d}+\frac{2|\theta|}{\pi}\int_{0}^\infty \frac{(dc-fa-eb+\tfrac{e^2a}{d})u^2+fc-\tfrac{f^2a}{d}}{d^2u^4+(2df-e^2)u^2+f^2}\D{u}+\mathcal{O}(\theta^2)\,,
\end{equation}
To prove this, we write
\begin{equation}
    \int_0^{2\pi}\D{u} \frac{a\sin^2 u+b\theta \sin u \cos u+c\theta^2}{d\sin^2 u+e\theta \sin u\cos u+f \theta^2}=2\pi\frac{a}{d}+\int_{-\pi}^{\pi}\D{u} \frac{(b-\tfrac{ea}{d})\theta \sin u \cos u+(c-\tfrac{fa}{d})\theta^2}{d\sin^2 u+e\theta \sin u\cos u+f \theta^2}\,.
\end{equation}
Then we decompose the integral into the intervals $[-\epsilon,\epsilon]$, $[\pi-\epsilon,\pi]$, $[-\pi,-\pi+\epsilon]$ and the rest denoted $I$, for some $\epsilon>0$. The limit $\theta\to 0$ can be taken directly on $I$, and using the change of variable $u\to-u$ one finds $I=\mathcal{O}(\theta^2)$. On $[-\epsilon,\epsilon]$, provided $\epsilon$ is small enough we can replace $\sin u$ by $u$ and $\cos u$ by $1$. Then we perform the change of variable $u=\theta v$ and take the limit $\theta\to 0$. With a similar process on the other intervals, we obtain \eqref{exp}.\\

Using (\ref{exp}), we find
\begin{equation}
    \sqrt{|\hat{\varphi}(k)|^2+|\hat{\psi}(k)|^2}=1-\gamma|\theta_k|+\mathcal{O}(\theta_k^2)\,,
\end{equation}
with after some algebra
\begin{equation}
    \gamma=\frac{4}{\pi}\int_0^\infty \D{u}\frac{|b|^2u^2+\tfrac{|b|^2}{|b|^2+|1-a|^2}}{(|b|^2+|1-a|^2)^2u^4+2[|b|^2+|1-a|^2-2(\Re(1-a))^2]u^2+1}\,,
\end{equation}
where the coefficients $a(k),b(k)$ are evaluated at $k=\pi/2$, and that we recall are defined as the coefficients of the matrix $\mathcal{M}_k$ in Eq. (17) of the main text, as in Eq. (8). Explicitly
\begin{equation}
a(\pi/2)=\sin^2(2x)e^{-4\lambda_c}-\cos^2(2x)e^{4ix}\,,\qquad b(\pi/2)=\sin(2x)\cos(2x)(e^{-4\lambda_c}+e^{4ix})\,.
\end{equation}
From \eqref{ent} we have then for $m\geq 2$
\begin{equation}
\begin{aligned}
     s_m(\lambda)&=\frac{2}{\pi}\int_{k_c}^{\pi/2}\D{k} H_m\left(\sqrt{|\hat{\varphi}(k)|^2+|\hat{\psi}(k)|^2} \right)\\
     &=-\frac{2 \gamma H_m'(1)}{\pi}\int_{k_c}^{\pi/2}\D{k}\theta_k+\mathcal{O}((\lambda_c-\lambda)^{3/2})\,.
\end{aligned}
\end{equation}
Using \eqref{thetak} and \eqref{kcexpand} we find
\begin{equation}
\begin{aligned}
      \int_{k_c}^{\pi/2}\D{k}\theta_k&=8\sin^2(2x)\cosh(2\lambda_c)(\lambda_c-\lambda)\int_0^1\D{y}\sqrt{1-y^2}+\mathcal{O}((\lambda_c-\lambda)^{3/2})\\
      &=2\pi\sin^2(2x)\cosh(2\lambda_c)(\lambda_c-\lambda)+\mathcal{O}((\lambda_c-\lambda)^{3/2})\,.
\end{aligned}
\end{equation}
Hence we obtain the following behaviour for $m\geq 2$
\begin{equation}
    s_m(\lambda)=2\gamma \sin^2(2x)\cosh(2\lambda_c)\frac{m}{m-1}(\lambda_c-\lambda)+\mathcal{O}((\lambda_c-\lambda)^{3/2})\,,
\end{equation}
and all the R\'enyi entropies have the critical exponent
\begin{equation}
    \nu=1\,.
\end{equation}
As for the von Neumann entropy, we have
\begin{equation}
    H_1(1-\epsilon)=\frac{\epsilon}{2}\left(1-\log \frac{\epsilon}{2}\right)+\mathcal{O}(\epsilon^2)\,.
\end{equation}
Hence it displays a marginally corrected critical behaviour as
\begin{equation}
    s_1(\lambda)=\gamma \sin^2(2x)\cosh(2\lambda_c)\frac{(1-2\log 2)}{2}(\lambda_c-\lambda)\log(\lambda_c-\lambda)+\mathcal{O}(\lambda_c-\lambda)\,.
\end{equation}
As for the Hartley entropy, we have
\begin{equation}
    s_0(\lambda)=\frac{2}{\pi}\int_{k_c}^{\pi/2}\D{k}\,,
\end{equation}
so
\begin{equation}
    s_0(\lambda)=\frac{4\sin(2x)}{\pi\sqrt{\tanh(2\lambda_c)}}\sqrt{\lambda_c-\lambda}+\mathcal{O}(\lambda_c-\lambda)\,,
\end{equation}
yielding the critical exponent
\begin{equation}
    \nu=\frac{1}{2}\,.
\end{equation}



\bibliography{bibliography}